\documentclass[reprint,nofootinbib,superscriptaddress,amsmath,amssymb,aps,prd]{revtex4-1}
\usepackage{amsmath,mathrsfs,amsbsy,color,graphicx,bm,amsthm,amsfonts}
\usepackage{bbm}
\usepackage{times}
\usepackage{units}
\usepackage{dcolumn}
\usepackage{subfig,graphicx}
\usepackage{epsfig}
\usepackage{epstopdf}
\usepackage[colorlinks,linkcolor=blue,anchorcolor=green,citecolor=blue,CJKbookmarks=True]{hyperref}
\DeclareMathSymbol{\shortminus}{\mathbin}{AMSa}{"39}
\usepackage{mathrsfs}
\usepackage{braket}
\usepackage{amssymb}
\usepackage{txfonts}
\usepackage{float}
\usepackage{enumitem}
\usepackage{multirow}
\usepackage{caption}
\usepackage[T1]{fontenc}

\DeclareUnicodeCharacter{0229}{\c{e}}



\begin{document}
	\title{Quantum coherence of continuous variables in the black hole quantum atmosphere}

	\author{Xiaofang Liu}
	\affiliation{Department of Physics, Key Laboratory of Low Dimensional Quantum Structures and Quantum Control of Ministry of Education, Hunan Research Center of the Basic Discipline for Quantum Effects and Quantum Technologies, Institute of Interdisciplinary Studies, and Synergetic Innovation Center for Quantum Effects and Applications, Hunan Normal
	University, Changsha, Hunan 410081, P. R. China}
	
	\author{Cuihong Wen}
	\email{cuihongwen@hunnu.edu.cn}
	\affiliation{College of information science and engineering, and Institute of Interdisciplinary Studies,  Hunan Normal University, Changsha, Hunan 410081, P. R. China}		
	
	\author{Jieci Wang}
	\email{jcwang@hunnu.edu.cn}
\affiliation{Department of Physics, Key Laboratory of Low Dimensional Quantum Structures and Quantum Control of Ministry of Education, Hunan Research Center of the Basic Discipline for Quantum Effects and Quantum Technologies, Institute of Interdisciplinary Studies, and Synergetic Innovation Center for Quantum Effects and Applications, Hunan Normal
	University, Changsha, Hunan 410081, P. R. China}
	
	\begin{abstract}	
Recently, the concept of quantum atmosphere has been introduced as a potential origin of Hawking quanta. This study investigates the properties of quantum fields by exploring the quantum coherence of a two-mode Gaussian state near a black hole, where Hawking quanta originate from the quantum atmosphere region. It is demonstrated that both physically accessible and inaccessible quantum coherence for continuous variable quantum states distinctly exhibit hallmark features of the quantum atmosphere. Specifically, the quantum coherence for these states varies continuously with changes in the normalized distance; it undergoes rapid decreases (or increases) just outside the event horizon before gradually stabilizing through subsequent increases (or decreases). This behavior contrasts with the behaviors of quantum coherence where originates solely from the black hole's event horizon. The quantum features of the fields distinctly reflect characteristics attributable to the quantum atmosphere, thereby deepening our understanding of the origins of Hawking radiation. We also find that the continuous variable coherence is highly dependent on both the squeezing parameter and field frequency of the prepared state; therefore, appropriately adjusting these values can enhance our ability to detect features within the quantum atmosphere. It is noteworthy to observe that quantum features of fields do not entirely dissipate in the quantum atmosphere region, indicating that tasks related to quantum information processing can still be executed there.			
	\end{abstract}
	
	\pacs{~}
	
	\maketitle

	\section{Introduction}
	
The discovery of Hawking radiation has opened a new chapter in the study of black holes \cite{hawking1975particle}, giving us a deeper understanding into the microscopic nature of gravity at the quantum level. Meantime, the insight also introduces some new theoretical problems in the semi-classical quantum field theory. Obviously, the most unresolved of which is the information loss paradox \cite{hawking1976breakdown,terashima2000entanglement,bombelli1986quantum}, namely, the thermal evaporation of the black hole does not match the unitary evolution specified by quantum mechanics. And, to better address the unitarity of the black hole and information paradox problem, it is critical to understand where Hawking radiation originates from. Hawking and Unruh argued that the Hawking quanta originate from the distance \textit{r} ($\Delta r=r-r_{H}\ll r_{H}$) near the black hole event horizon $ r_{H}$ \cite{hawking1975particle,unruh1977origin},  which represents the prevailing view. However, Giddings recently questioned this idea \cite{giddings2016hawking}, claiming that Hawking quanta do not originate at the event horizon but rather in the quantum atmosphere region $r_A$ ($\Delta r=r_A-r_H\sim r_H$) outside the event horizon, based on investigations into the total emissivity and stress tensor of Hawking radiation. Giddings' frameworks were further backed up by the Stefan-Boltzmann law inferring the size of the radiator \cite{giddings2016hawking}. Subsequently, several related studies have corroborated this finding by analyzing the Schwinger effect and stress-energy tensor \cite{dey2017black,dey2019black,ong2020quantum}, as well as investigating Hawking radiation spectra in arbitrary dimensions \cite{hod2016hawking} and local temperature within a semi-classical framework \cite{eune2019test}. 
	
The quantum atmosphere represents not only an intriguing new concept regarding the origin of Hawking radiation but also signifies a substantial transformation in the physical properties surrounding black holes. Analyzing this phenomenon from the perspective of relativistic quantum information \cite{fuentes2005alice,blasco2015violation,chkecinska2015communication,richter2015degradation,perche2023role,garttner2023general,liu2023entanglement,liu2022gravity,tian2023direct,li2023quantum,calmet2017transformation,anastopoulos2023quantum,shamsi2020analysis,tjoa2022quantum,perche2023role,camblong2024entanglement,harikrishnan2022accessible,liu2023quantum,liu2024optimal}  can provide deeper insights into both the theory of quantum information and related physical phenomena within the framework of general relativity. That is, focusing further on quantum correlations around the black hole may enhance our understanding of the quantum atmosphere as well as the microscopic nature of gravity at the quantum level  \cite{kaczmarek2024signatures,kaczmarek2024coherence}. Quantum coherence \cite{baumgratz2014quantifying,streltsov2015measuring}  is one of the fundamental concepts in quantum physics, embodying the superposition capability of quantum states and serving as an essential characteristic of quantum systems. Similar to quantum entanglement, it also represents a crucial quantum resource that applies to quantum information processing tasks, including quantum information storage \cite{gundougan2012quantum,li2020quantum}, quantum computing and quantum metrology.  As a result, an in-depth understanding of quantum coherence in the quantum atmosphere is of great significance for the growth of modern quantum technologies.	
	
In this paper, we investigate the dynamics of continuous variable quantum states \cite{weedbrook2012gaussian,grochowski2017effect,dkebski2018multimode} for scalar fields within the region of the quantum atmosphere and explore new features of quantum coherence of continuous variable states \cite{xu2016quantifying} in this context. Building upon Gidding's argument, we conduct a second quantization of the scalar field in the quantum atmosphere region. Subsequently, we obtain the relationship between the local Hartle--Hawking temperature and the effective radius of the black hole quantum atmosphere. Within this framework, we demonstrate that the evolution of the quantum state of the field and the dynamics of continuous variable quantum coherence are closely contingent on the effective temperature induced by the black hole atmosphere. Interestingly, quantum features of fields do not entirely dissipate in the quantum atmosphere region, showing that quantum information processing tasks can still be performed there. The quantum coherence for continuous variable quantum states exhibits features of the quantum atmosphere, thereby contributing to a deeper understanding of the origin of Hawking radiation.

This paper is structured as follows: In Sec. \ref{Sec.2}, we outline the vacuum structure of the scalar field within the quantum atmosphere of the Schwarzschild spacetime. In Sec. \ref{Sec.3}, we analyze the features of quantum coherence for the initially correlated and uncorrelated modes under the influence of the black hole's quantum atmosphere. In Sec. \ref{Sec.4}, we summarize the paper.

	\section{Vacuum structure of the scalar field in the quantum atmospheric region }\label{Sec.2}
	In this section, we present the second quantization of the scalar field and its vacuum structure within the quantum atmosphere region of a Schwarzschild black hole.
	\begin{align}\label{eq1}ds^{2}&=-\left(1-\frac{2M}{r}\right)dt^{2}+\left(1-\frac{2M}{r}\right)^{-1}dr^{2}\notag\\&+r^{2}(d\theta^{2}+\sin^{2}\theta d\varphi^{2}),\end{align}
	where \textit{M} represents the mass of the black hole.
	
	The dynamics of the massless scalar field is given by the Klein-Gordon equation \cite{birrell1984quantum}
	\begin{equation}
		\frac{1}{\sqrt{-g}}\frac{\partial}{\partial x^{\mu}}\left(\sqrt{-g}g^{\mu\nu}\frac{\partial\phi }{\partial x^{\nu}}\right)=0.\label{equation2}
	\end{equation}
	By solving Eq. (\ref{equation2})  near the event horizon of the black hole, one obtains a series of outgoing modes  both in the inner region and the quantum atmospheric region of the black hole. Then the scalar field $\Phi$ in the quantum atmosphere region can be expanded as \cite{birrell1984quantum}
	\begin{align}
		\Phi=\sum_{lm}\int d\omega\Big[b_{\mathrm{in},\omega lm}\phi _{out,\omega lm}(r<r_{A})+b_{\mathrm{in},\omega lm}^{\dagger}\phi_{out,\omega lm}^{*}(r<r_{A})\notag
		\\+b_{out,\omega lm}\phi _{out,\omega lm}(r>r_{A})+b_{out,\omega lm}^{\dagger}\phi _{out,\omega lm}^{*}(r>r_{A})\Big],\label{equation8}
	\end{align}
    where $b_{\mathrm{in},\omega lm}^\dagger $ and $b_{\mathrm{in},\omega lm}$ are creation and annihilation operators acting on the vacuum inside the Schwarzschild black hole, $b_{out,\omega lm}^{\dagger}$ and $b_{out,\omega lm }$ are creation and annihilation operators acting on the vacuum state in the quantum atmospheric region. And  $r_A$  is the effective radius of the Schwarzschild black hole quantum atmosphere \cite{hod2016hawking}
	\begin{equation}
		r_{\mathrm A}=\Big[\frac{256\pi^{5}}{3\zeta(4)} \bar{P}_{\mathrm B\mathrm H}\Big]^{\frac{1}{2}}\times r_{\mathrm H},
	\end{equation}
where  $\bar{P}_{\mathrm{BH}}\equiv P_{\mathrm{BH}}\times\frac{r_{\mathrm{H}}^{2}}{\hbar}$, and $P_{\mathrm{BH}}$ is the semi-classical Hawking radiation power.\\
By performing an analytic continuation of the Schwarzschild modes with Kruskal-like coordinates, we obtain a new set of completed field modes. Naturally, we can expand the scalar field $\Phi$ by employing the  Kruskal-like modes
	\begin{align}
		\Phi&=\sum_{lm}\int d\omega[2\sinh(4\pi\omega M)]^{-1/2}[a_{I,\omega lm}\phi_{I,\omega lm}\notag
		\\&+a_{I,\omega lm}^{\dagger}\phi_{I,\omega lm}^{*}+a_{II,\omega lm}\phi_{II,\omega lm}+a_{II,\omega lm}^{\dagger}\phi_{II,\omega lm}^{*}],\label{equation16}
	\end{align}
	where $a_{I,\omega lm} $ and $a_{II,\omega lm} $  represent the annihilation operator acting on the Kruskal vacuo.
Then one can compute the Bogoliubov relationships for the annihilation and creation operators acting on different vacuo \cite{wang2009entanglement}
	\begin{align}
		a_{I,\omega lm}=b_{out,\omega lm}\cosh u-b_{in,\omega lm}^{\dagger}\sinh u,
		\\a_{I,\omega lm}^{\dagger}=b_{out,\omega lm}^{\dagger}\cosh u-b_{in,\omega lm}\sinh u,
	\end{align}
  where $\cosh u = \frac{1}{\sqrt{1-e^{-\frac{\omega}{T_{HH}}}}}$, and $T_{HH}$ is the effective temperature in the quantum atmosphere region \cite{eune2019test}
	\begin{align}
		&T_{HH}=T_H\sqrt{1-\frac{r_h}r}\notag
		\\&\sqrt{1+2\frac{r_h}r+\left(\frac{r_h}r\right)^2\left(9+4 D_{HH}+36\ln\left(\frac{r_h}r\right)\right)},\label{equation29}
	\end{align}
	where $T_H = \frac{1}{4\pi r_h}$, and $D_{HH}$ is the undetermined constant of the stress tensor in the Hartle--Hawking vacuum, i.e., the Hartle--Hawking parameter. And, as $D_{HH}$ increases, the peaks of the local
	temperatures in equilibrium lie in $1.43r_h\lesssim r_{\mathrm{peak}}<1.5r_h$. It is also worth noting that at the black hole event horizon $T_{HH}=0$, and when $r\rightarrow\infty $ it approaches the standard  Hawking temperature of the Schwarzschild black hole.

	After normalizing the state vectors, it is found that the Kruskal vacuum state can be represented by an entangled two-mode squeezed state \cite{adesso2007continuous}
	\begin{equation}
		|0\rangle_{K}=\frac{1}{\cosh u}\sum_{n=0}^{\infty}\tanh^{n}u |n\rangle_{in}|n\rangle_{out},\label{equation20}
	\end{equation}
	and the single-particle excited state is expressed as \cite{fuentes2005alice,birrell1984quantum} 
	\begin{equation}
		|1\rangle_{K}=\frac{1}{\cosh^{2}u}\sum_{n=0}^{\infty}\tanh^{n}u\sqrt{(n+1)}\left|\left(n+1\right)\right\rangle_{in}\otimes\left|n\right\rangle_{out},
	\end{equation}
	where $|n\rangle_{in}$ and $|n\rangle_{out}$ are the excited states of local field modes inside the  event horizon and in  the  quantum atmosphere region, respectively.
	
	\section{Behavior of quantum coherence  in the quantum atmospheric region}\label{Sec.3}
	\subsection{Coherence  between the initially correlated modes}
	To explore the properties of field modes under the different origins of Hawking quanta, in this subsection, we analyze the quantum coherence between the initially correlated modes under the influence of quantum atmosphere. We assume that the observers Alice and Bob share a two-mode squeezed initial state and that Alice places on an asymptotically flat area, while Bob, after a period of free fall, hovers at the distance \textit{r} from the center of the black hole.
	The two-mode squeezed  state is expressed by the covariance matrix \cite{adesso2007continuous}
	\begin{equation}
		\varrho_{AB}(s)=\left(\begin{array}{cc}\cosh(2s)I_2&\sinh(2s)Z_2\\[0.3em]\sinh(2s)Z_2&\cosh(2s)I_2\end{array}\right).
	\end{equation}
	where $\left.I_2=\left(\begin{array}{cc}1&0\\0&1\end{array}\right.\right)$, $\left.Z_2=\left(\begin{array}{cc}1&0\\0&-1\end{array}\right.\right)$, and \textit{s} is the squeezing parameter. At the same time, Alice and Bob are equipped with detectors A and B, respectively. Naturally, according to Giddings' argument, the detector B at radius \textit{r} will experience thermal radiation emitted by the quantum atmosphere region. It is found that the influence of thermal radiation can be expressed in terms of the  two-mode squeezed transformation of Eq. (\ref{equation20}), which is described in phase space by the  symplectic operator as
	\begin{equation}
		S_{B,\bar B}(u)=\left(\begin{array}{cc}\cosh(u)I_2&\sinh(u)Z_2\\[0.3em]\sinh(u)Z_2&\cosh(u)I_2\end{array}\right).
	\end{equation}
By applying the two-mode squeezed operation, the mode \textit{B} is mapped to the mode \textit{B} and $\bar{B}$, which are in the quantum atmosphere region and the regions inside the event horizon, respectively. The system is divided into three parts: the subsystem \textit{A} described by Alice, the subsystem \textit{B} described by Bob of the region outside the event horizon, and the subsystem $\bar{B}$ described by the virtual observer anti-Bob, whose causal access is limited to the region inside the horizon. The total covariance matrix after the effect of the two-mode squeezed operation is \cite{adesso2007continuous}
	\begin{equation}
		\varrho_{AB\bar{B}}(s,u)=\begin{bmatrix}I_{A}\oplus S_{B,\bar{B}}(u)\end{bmatrix}\begin{bmatrix}\varrho_{AB}(s)\oplus I_{\bar{B}}\end{bmatrix}\begin{bmatrix}I_{A}\oplus S_{B,\bar{B}}(u)\end{bmatrix},
	\end{equation}
	where $I_{A}\oplus S_{B,\bar{B}}(u)$ signifies the two-mode squeezed transformation between the subsystems \textit{B} and $\bar{B}$, and $\varrho_{AB}(s)\oplus I_{\bar{B}}$ denotes the initial covariance matrix for the entire system.
	Since the regions inside and outside the event horizon are causally disconnected, Alice and Bob in the region outside the event horizon cannot access the mode $\bar{B}$ inside the event horizon.  Therefore, we obtain the covariance matrix between Alice and Bob by tracing out the mode $\bar{B}$
	\begin{equation}
		\varrho_{AB}(s,u)=\begin{pmatrix}\mathcal{A}_{AB}&\mathcal{C}_{AB}\\\mathcal{C}_{AB}^\top&\mathcal{B}_{AB}\end{pmatrix},\label{equation25}
	\end{equation}
	where
	$$\mathcal{A}_{AB}=\cosh(2s) I_{2},$$
	$$C_{AB}=\cosh u\sinh(2s)Z_{2},$$
	$$\mathcal{B}_{AB}=[\cosh^2u\cosh2s+\sinh^2u]I_2.$$
	
Quantum coherence for a continuous variables state  is defined by $C(\rho)=\mathrm{inf}_{\delta}\{S(\rho||\delta)\}$, where $S(\rho||\delta)=\mathrm{tr}(\rho\mathrm{log}_{2}\rho)-\mathrm{tr}(\rho\mathrm{log}_{2}\delta)$ is the relative entropy, $\mathbf{\delta}$ is an incoherent continuous variables state. For a two-mode continuous variable state, the quantum coherence can be  expressed as \cite{xu2016quantifying,wu2019quantum,xiao2022generation}
	\begin{equation}
		C(\rho)=-S(\rho)+\sum_{i=1}^{2} \left[\left(\overline{n}_{i}+1\right) \log_{2} \left(\overline{n}_{i}+1\right)- \overline{n}_{i} \log_{2}\overline{n}_{i}\right],\label{equation26}
	\end{equation}
	where $S(\rho)$ is the von Neumann entropy, which can be written as the symsymtic eigenvalues  $\nu_\pm$ of the continuous variable state \cite{holevo1999capacity}
	\begin{equation}
		S(\rho)=f(\nu_+)+f(\nu_-),
	\end{equation}
	where $f(\nu)=\frac{\nu+1}{2}\log_{2}\frac{\nu+1}{2}-\frac{\nu-1}{2}\log_{2}\frac{\nu-1}{2}$, and $\nu_\pm=\sqrt{\frac{\Delta\pm\sqrt{\Delta^2-4\det\mathbf{\varrho_{AB}}}}2}$, $\Delta:=\det\mathcal{A}+\det\mathcal{B}+2\det\mathcal{C} $ \cite{weedbrook2012gaussian}. The average number of particles $\bar{n}_i$ of the mode \textit{i} is expressed as
	\begin{equation}
		\overline{n}_i=\frac{1}{4}(\sigma_{11}^i+\sigma_{22}^i+[d_1^i]^2+[d_2^i]^2-2),
	\end{equation}
	where $\sigma_{jj}^i$ are the elements of the reduced covariance matrix, $d_j^i$ is the \textit{j} first statistical moment of the \textit{i} mode.

	\begin{figure}[H]
		\centering
		\includegraphics[scale=0.28]{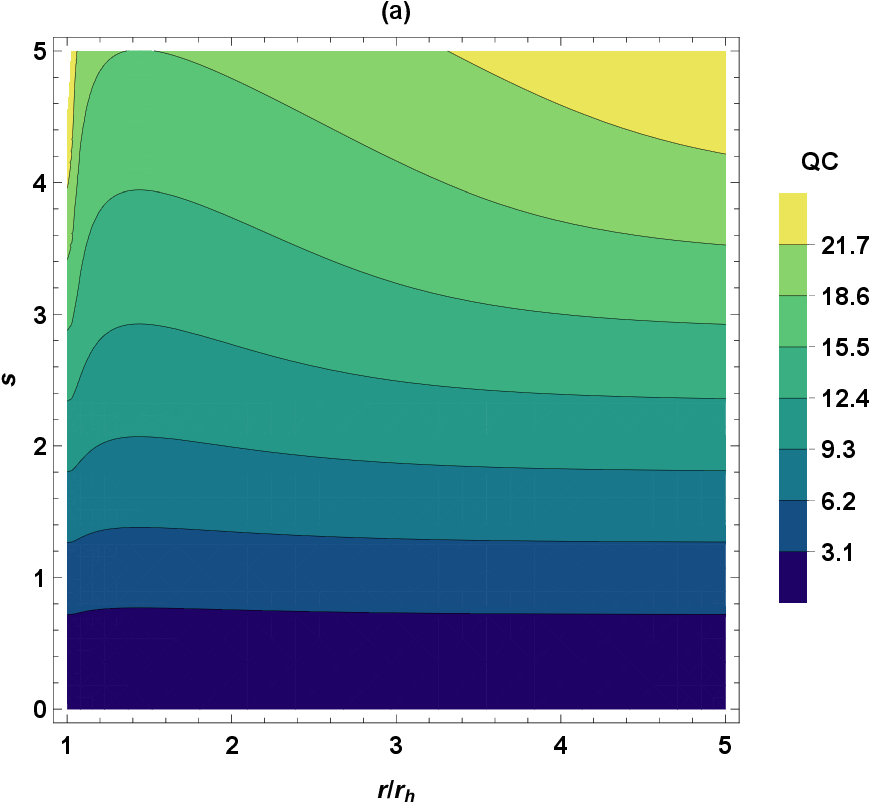}
		\includegraphics[scale=0.28]{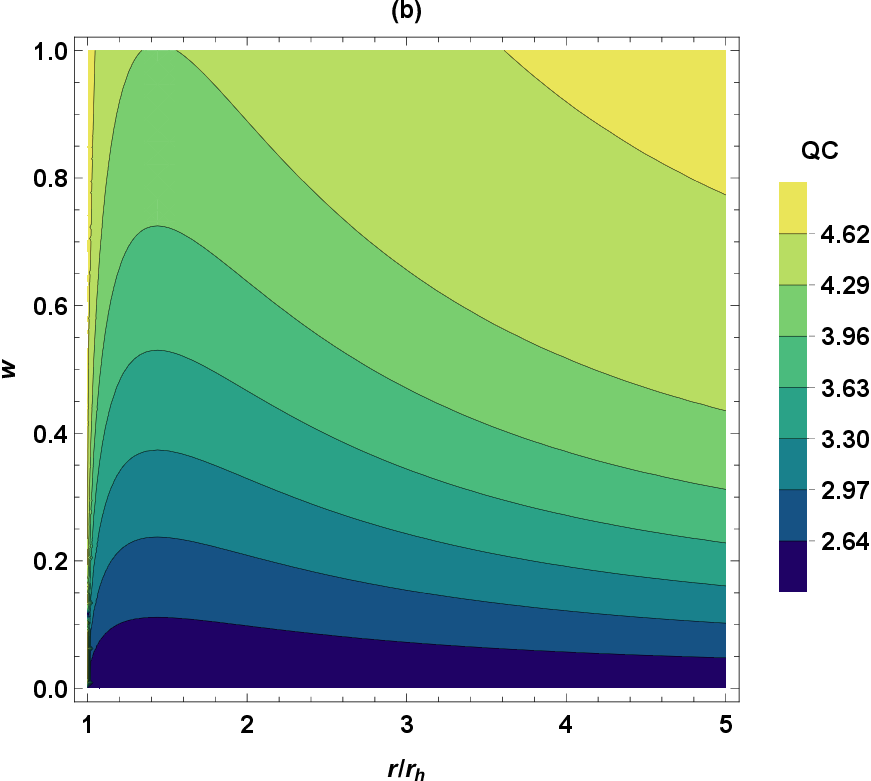}
		\includegraphics[scale=0.5]{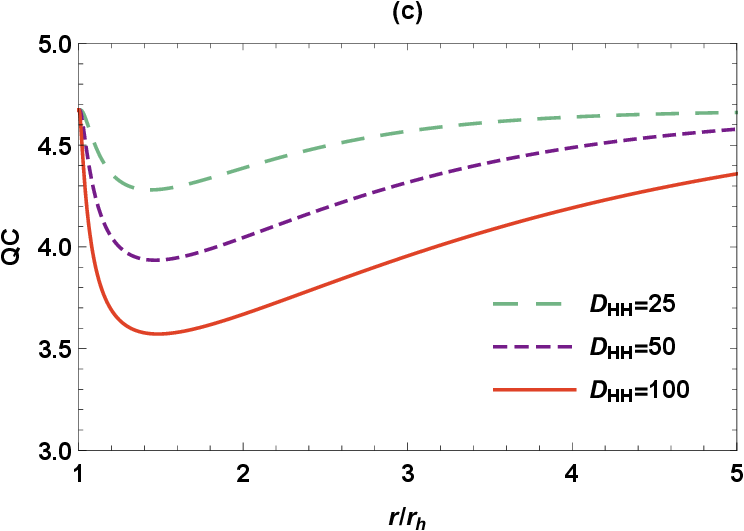}
		\caption{(a) Quantum coherence between initially correlated modes as a function of the initial squeezing parameter \textit{s} and the normalized distance ($r/r_{h}$). The frequency is set to \textit{w} = 1, and the Hartle--Hawking parameter is set to $D_{HH}=25$.
		(b) The coherence varies with the frequency \textit{w} and the normalized distance ($r/r_{h}$). The initial squeezing parameter is set to \textit{s} = 1, and the Hartle--Hawking parameter is set to $D_{HH}=25$.	(c) The physically accessible coherence for different Hartle--Hawking parameters ($D_{HH}$). The remaining parameters are fixed at \textit{s} = \textit{w} = 1.}\label{fig1}
	\end{figure}\noindent		
	
By utilizing the covariance matrix  of Eq. (\ref{equation25}), in conjunction with Eqs. (\ref{equation29}) and (\ref{equation26}), we can assess the impact of the quantum atmosphere on quantum coherence for initially correlated modes.
	
In Fig. \ref{fig1}(a), we demonstrate the evolution of physically accessible quantum coherence as a function of the initial squeezing parameter \textit{s} and the normalized distance ($r/r_{h}$). The physically accessible quantum coherence attains its maximum at $r/r_{h} = 1$ and exhibits a non-monotonic trend as the position of detector B varies, increasing to a stable value after $r\approx1.43r_{h}$ reaches a minimum, which contrasts with the monotonic behavior of quantum coherence of continuous variables observed in Hawking radiation originating from the event horizon \cite{wu2019quantum}. Based on the relationship between the local temperature and the normalized distance, for given $r_{h}$, the radial \textit{r} is a function of the variation in local temperature, transitioning from zero to a maximum value before subsequently decreasing to the Hawking temperature. 
	
At $r\approx1.43r_{h}$, the thermal radiation from the quantum atmosphere peaks, significantly degrading the physically accessible quantum coherence. However, with a sufficiently large initial squeezing parameter \textit{s}, Alice and Bob's quantum coherence remains intact even in regions of the highest local temperature, suggesting that adequate initial resources to carry out quantum information processing tasks in the quantum atmosphere. Notably, when \textit{s} is very small, the physically accessible quantum coherence shows little change in the quantum atmosphere; conversely, a larger initial two-mode squeezing parameter reveals a clear trend in coherence, indicating its substantial impact on the characteristics of the quantum atmosphere.
	
Fig. \ref{fig1}(b) shows the variation of quantum coherence with frequency \textit{w} and the normalized distance ($r/r_{h}$). The physically accessible quantum coherence of the state $\rho_{A{B}}$ is zero regardless of the frequency value when $r/r_{h}=1$,  indicating that there is no coherence generated at this point. As the normalized distance varies, the frequency \textit{w} significantly influences the trend of physically accessible quantum coherence within the quantum atmosphere region. Moreover, it is evident that low-frequency scalar fields facilitate the detection of features in the quantum atmosphere, suggesting that scalar field frequency positively impacts the characterization of this environment.

Fig. \ref{fig1}(c) illustrates the variation of quantum coherence with normalized distance when the Hartle--Hawking parameter  is set to various fixed values. The graph indicates that the minimum of physically accessible quantum coherence occurs at $r\approx1.43r_{h}$, independent of the Hartle--Hawking parameter. Notably, the value of the Hartle--Hawking parameter is proportional to the loss of initial quantum coherence. According to Eq. \ref{equation29}, increasing the Hartle--Hawking parameter while keeping detector B stationary results in a rise in local temperature within the quantum atmosphere, leading to greater degradation of quantum coherence. In other words,  the Hartle--Hawking parameter plays a key role in the dynamics of quantum coherence in the quantum atmosphere.

\subsection{Quantum coherence between the causal disconnected modes}
	
In this subsection, we analyze the impact of the quantum atmosphere on the the quantum coherence between the causal disconnected modes. The covariance matrix  between modes \textit{B} and $ \bar{B} $ is derived by tracing out the mode \textit{A}
	\begin{equation}
		\varrho_{B\bar B}(s,u)=\begin{pmatrix}\mathcal{A}_{B\bar B}&\mathcal{C}_{B\bar B}\\\mathcal{C}_{B\bar B}^\top&\mathcal{B}_{B\bar B}\end{pmatrix},\label{equation30}
	\end{equation}
	where
	\[\mathcal{A}_{B\bar{B}}=\begin{bmatrix}\cosh(2s)\cosh^2(u)+\sinh^2(u)\end{bmatrix}I_2,\]
	\[\mathcal{B}_{B\bar B}=\begin{bmatrix}\cosh^2(u)+\cosh(2s)\sinh^2(u)\end{bmatrix}I_2,\]
	\[\mathcal{C}_{B\bar{B}}=\begin{bmatrix}\cosh^2(s)\sinh(2u)\end{bmatrix}Z_2.\]
	\begin{figure}[H]
		\centering
		\includegraphics[scale=0.3]{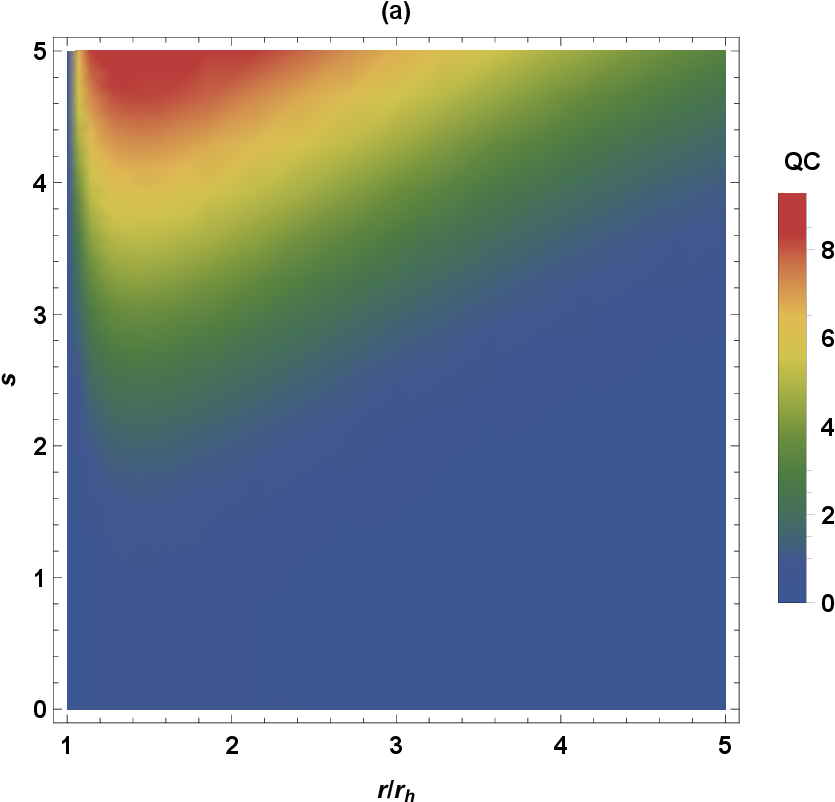}
		\includegraphics[scale=0.3]{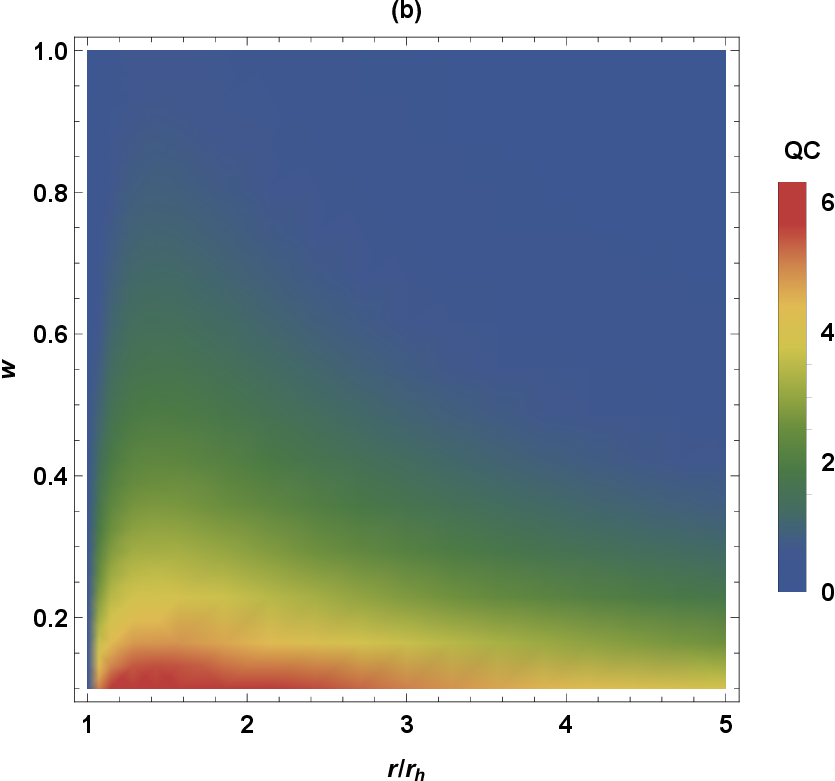}
		\includegraphics[scale=0.5]{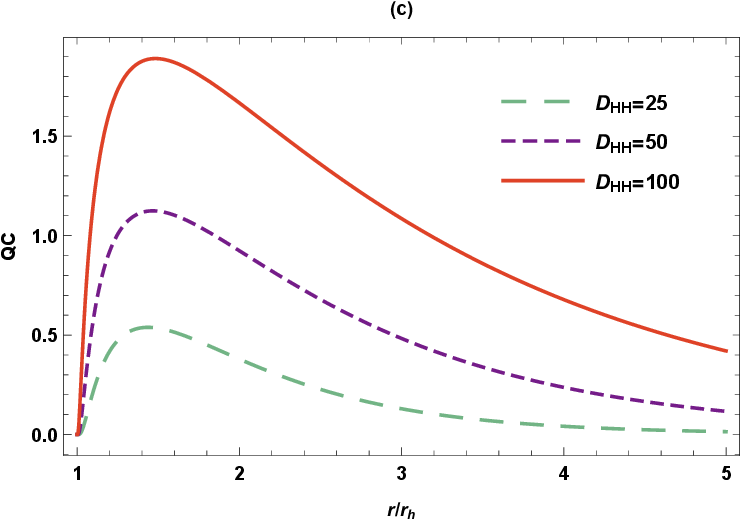}
		\caption{(a) Density plot of quantum coherence between Bob and anti-Bob as a function of the initial squeezing parameter \textit{s} and the normalized distance ($r/r_{h}$). The frequency \textit{w} is fixed at 1 and the Hartle--Hawking parameter $D_{HH}$ is fixed at 25. (b) Plot of physically inaccessible quantum coherence as a function of the frequency \textit{w} and the normalized distance ($r/r_{h}$). The initial squeezing parameter \textit{s} is fixed at 1, and the Hartle--Hawking parameter $D_{HH}$ is fixed at 25. (c) Two-dimensional plots showing the quantum coherence of the initially uncorrelated modes as a function of the normalized distance ($r/r_{h}$) for various values of the Hartle--Hawking parameter $D_{HH}$ with the initial squeezing parameter \textit{s} and the frequency \textit{w} fixed at 1.}\label{fig2}
	\end{figure}
Similarly, by employing the covariance matrix of Eq. (\ref{equation30}),  combined with Eqs. (\ref{equation29}) and (\ref{equation26}),  we can also compute the results for the effect of the quantum atmosphere on the physically inaccessible quantum coherence. From Fig. \ref{fig2}(a), it is evident that the physically inaccessible quantum coherence between modes \textit{B} and  $ \bar{B} $ initially exhibits a sharp increasing trend as the normalized distance ($ r/r_{h} $)  begins to rise, indicating that the local temperature in the quantum atmosphere generates quantum coherence between the initially uncorrelated modes. In other words, the presence of non-zero coherence suggests that quantum resources are, in principle, available within the region between the event horizon and the quantum atmosphere, even if the practical implementation of information processing tasks is not feasible due to the causal disconnection.  Furthermore, after reaching its maximum at $r\approx1.43r_{h}$, the physically inaccessible quantum coherence subsequently returns to a stable value.
	
	Fig. \ref{fig2}(b) illustrates the variations of quantum coherence between uncorrelated modes as a function of the field frequency \textit{w} and normalized distance $r/r_{h}$. We similarly observe that changes in quantum coherence are more pronounced at low-frequency scalar fields for the Gaussian state  $\rho_{B\bar{B}}$. This suggests that, while physically accessible and inaccessible quantum coherence are plotted opposite for various scalar field frequencies, they exhibit identical characteristics of the quantum atmosphere as the field frequency varies. Fig. \ref{fig2}(c) presents a nonmonotonic two-dimensional map of distance-dependent physically inaccessible quantum coherence. As previously noted, the peak of this plot occurs at $r\approx1.43r_{h}$. Additionally, it is observed that initial quantum coherence for the state $\rho_{B\bar{B}}$ is more readily obtained with an increased Hartle--Hawking parameter, further indicating that within the quantum atmosphere region of the black hole, the Hartle--Hawking parameter plays a crucial role in the dynamics of quantum coherence. 
	
	It is worth noting that although our analysis spans the entire physically accessible region \((r > r_h)\), our primary aim is to characterize the quantum atmosphere by examining the evolution of quantum coherence. We treat the radial distance \(r\) as the independent variable. We find that the quantum coherence reaches an extremum at \(r \approx 1.43\,r_h\). This location coincides with the peak of the local Hartle--Hawking temperature, thereby revealing the physical features of the quantum atmosphere at the effective radius \(r_A\).

	\section{Conclusion}\label{Sec.4}
In this manuscript, we investigate the dynamics of continuous variable quantum states for scalar fields within the quantum atmosphere region and explore different features of coherence in this context. We demonstrate that both field state evolution and the dynamics of continuous variable quantum coherence are closely linked to effective temperatures induced by black hole atmospheres. We find that the Hartle--Hawking parameter plays a crucial role in the dynamics of quantum coherence in the black hole quantum atmosphere region. Furthermore, it is shown that coherence in continuous variable states distinctly exhibits features attributable to the quantum atmosphere, thereby enhancing our understanding of the origins of Hawking radiation. The continuous variable coherence is highly dependent on both the squeezing parameter and the field frequency of the prepared state; therefore, adjusting these values appropriately can enhance our ability to detect features within the quantum atmosphere. It is also shown that quantum correlations do not entirely vanish in the quantum atmosphere region, indicating that certain quantum information processing tasks can still be performed there, which is highly significant. The present work not only sheds light on how quantum resources behave under the influence of thermal radiation produced in the quantum atmosphere region but also characterizes the quantum atmosphere, therefore offering valuable insights for exploring the various origins of Hawking radiation within the framework of relativistic quantum information.
\\

	\acknowledgments
	This work was supported by the National Natural Science Foundation of China under Grants  No.  No.12475051 and No.12374408; the science and technology innovation Program of Hunan Province under grant No.2024RC1050; and the innovative research group of Hunan Province under Grant No.2024JJ1006.
	
	\bibliographystyle{apsrev4-1}
	
	%

\end{document}